\documentclass[preprint,superscriptaddress, amsmath,amssymb, aps]{revtex4-1}

\usepackage{graphicx}
\usepackage{dcolumn}
\usepackage{bm}
\usepackage[normalem]{ulem}
\usepackage{soul}
\usepackage[utf8]{inputenc}
\usepackage[T1]{fontenc}
\usepackage{comment}

\newcommand{\degree}{$^{\circ}$}

\newcommand{\figuretitle}[1]{{\textbf{#1} \par \medskip}}
\usepackage{cases}
\usepackage{amsmath}

\usepackage{hyperref}

\usepackage{xcolor}

\DeclareUnicodeCharacter{00A0}{~}

\begin{document}

\title{Direct observation of N{\'e}el-type skyrmions and domain walls in a ferrimagnetic DyCo$_3$ thin film} 

\author{Chen Luo}
\affiliation{Helmholtz-Zentrum-Berlin f\"{u}r Materialen und Energie, Albert-Einstein-Stra{\ss}e 15, 12489 Berlin, Germany}
\affiliation{Institute of Experimental Physics of Functional Spin Systems, Technical University Munich, James-Franck-Stra{\ss}e 1, 85748 Garching b. M\"{u}nchen, Germany}

\author{Kai Chen}
\email{kaichen2021@ustc.edu.cn}
\affiliation{Helmholtz-Zentrum-Berlin f\"{u}r Materialen und Energie, Albert-Einstein-Stra{\ss}e 15, 12489 Berlin, Germany}
\affiliation{National Synchrotron Radiation Laboratory, University of Science and Technology of China, Hefei, Anhui 230029, China}

\author{Victor Ukleev}
\affiliation{Helmholtz-Zentrum-Berlin f\"{u}r Materialen und Energie, Albert-Einstein-Stra{\ss}e 15, 12489 Berlin, Germany}

\author{Sebastian Wintz}
\affiliation{Helmholtz-Zentrum-Berlin f\"{u}r Materialen und Energie, Albert-Einstein-Stra{\ss}e 15, 12489 Berlin, Germany}
\affiliation{Max-Planck-Institut f\"{u}r Intelligente Systeme, 70569 Stuttgart, Germany}

\author{Markus Weigand}
\affiliation{Helmholtz-Zentrum-Berlin f\"{u}r Materialen und Energie, Albert-Einstein-Stra{\ss}e 15, 12489 Berlin, Germany}
\affiliation{Max-Planck-Institut f\"{u}r Intelligente Systeme, 70569 Stuttgart, Germany}
\author{Radu-Marius Abrudan}
\affiliation{Helmholtz-Zentrum-Berlin f\"{u}r Materialen und Energie, Albert-Einstein-Stra{\ss}e 15, 12489 Berlin, Germany}

\author{Karel Proke\v{s}}
\affiliation{Helmholtz-Zentrum-Berlin f\"{u}r Materialen und Energie, Hahn-Meitner Platz 1, D-14109 Berlin, Germany}

\author{Florin Radu}
 \email{florin.radu@helmholtz-berlin.de}
\affiliation{Helmholtz-Zentrum-Berlin f\"{u}r Materialen und Energie, Albert-Einstein-Stra{\ss}e 15, 12489 Berlin, Germany}

\date{\today}

\begin{abstract}

Isolated magnetic skyrmions are stable, topologically protected spin textures that are at the forefront of research interests today due to their potential applications  in information technology. 
A distinct class of skyrmion hosts are rare earth - transition metal (RE-TM) ferrimagnetic materials. 
To date, the nature and the control of basic traits of skyrmions in these materials are not fully understood. We show that for an archetypal ferrimagnetic material 
DyCo$_3$
that exhibits a strong perpendicular anisotropy, the ferrimagnetic skyrmion size can be tuned by an external magnetic field.  Moreover, by taking advantage of the high spatial resolution of scanning transmission X-ray microscopy (STXM) and utilizing a large x-ray magnetic linear dichroism (XMLD) contrast that occurs naturally at the RE resonant edges, we resolve the nature of the magnetic domain walls of ferrimagnetic skyrmions. We demonstrate that through this method one can easily discriminate between Bloch and N\'{e}el type domain walls for each individual skyrmion. For all isolated ferrimagnetic skyrmions, we observe that the domain walls are of N\'{e}el-type. This key information is corroborated  with results of  micromagnetic simulations and allows us to conclude on the nature of the Dzyaloshinskii–Moriya
interaction (DMI) which concurs to the stabilisation of skyrmions in this  ferrimagnetic system. Establishing that an intrinsic DMI occurs in RE-TM materials will also be beneficial towards a deeper understanding of chiral spin texture control in ferrimagnetic materials.
\end{abstract}
\maketitle

\section*{INTRODUCTION}

Magnetic skyrmions are stable nanoscale whirls of magnetic spin textures \cite{skyrme1962unified,skyrme1962unified,roessler2006spontaneous,yu2010real,munzer2010skyrmion,sampaio2013nucleation,fert2017magnetic}.
Due to their topological stability, small size at the nanometer scale and controlled mobility under low current densities, skyrmions hold the promise to impact significantly   next-generation information storage technology~\cite{parkin2008magnetic,fert2013skyrmions,parkin2015memory,zhang2015skyrmion,tomasello2014strategy,kang2016voltage,tomasello2017performance,Zhang2020-re,hoffmann2021skyrmion,one:2022}. 
Initially introduced  in nuclear physics as soliton solutions of non-linear field equations~\cite{skyrme1961non,skyrme1962unified}, they hold now a distinct  place in   solid state physics as well, following their theoretical prediction~\cite{Bogdanov1989} and experimental observation~\cite{doi:10.1126/science.1166767}. Broken inversion symmetry that is characteristic  to certain crystal structures induces a non-collinear coupling mechanism that contributes as an asymmetric term in the Hamiltonian describing the resulting magnetically chiral ground states~\cite{bogdanov1994thermodynamically}.  Under certain conditions, these chiral states concur in forming magnetic skyrmions as observed in archetypal cubic chiral crystals that exhibit a bulk Dzyaloshinskii–Moriya
interaction (DMI)~\cite{doi:10.1126/science.1166767,doi:10.1126/science.1214143,neubauer2009topological,onose2012observation,pollath2019ferromagnetic,ukleevSTAM2022}. Moreover, symmetry breaking along with spin-orbit coupling present at 
magnetic interfaces lead to a weak interfacial DMI that contributes to the stabilisation of skyrmions observed in  thin films~\cite{yu2011near,heinze2011spontaneous,bode2007chiral,yu2016room,jiang2015blowing}  and  multilayers~\cite{romming2013writing,moreau2016additive,woo2016observation,woo2017spin,hrabec2017current,boulle2016room}.

Skyrmion  lattices that occur in single crystals fill a small pocket in the phase diagram for temperatures that typically extends over few Kelvin~\cite{doi:10.1126/science.1166767,PhysRevB.88.195137}. This is detrimental to applications which  require stability over a broad range around room temperature. The temperature pocket can be eventually extended by reducing the dimensionality of the structures or by engineering the interfacial DMI of ferromagnetic thin films and multilayers~\cite{ma2022}. Yet,  caused by the skyrmion Hall effect~\cite{Jiang2017,Litzius2017}, the trajectories of these ferromagnetic topological units in devices are not straight, being deflected away by the Lorentz forces. To overcome this limitation, ferrimagnetic materials are offering an advantage due to a versatile tunability of their magnetic properties.
 
Rare-earth-transition-metals (RE-TM) ferrimagnetic (FiM) alloys consist two anitferromagnetically coupled sublattices. At the compensation temperature (T$_{comp}$), the  magnetizations of both sub-lattices are equal, leading to a vanishing net magnetization, just as for an antiferromagnet. By the choice of the elemental composition and through  temperature variation, their  magnetic properties, including T$_{comp}$, net magnetization and magnetic anisotropy,  can be easily engineered, which makes FiM materials advantageous for spintronics devices \cite{RADU2018267,kim2022ferrimagnetic,sala2022}. By selecting the RE element, two classes of FiM can be distinguished, namely Gd-base FiM alloys that exhibit a weak perpendicular magnetic anisotropy (PMA) due to the vanishing orbital moment of the RE, and Dy, Tb, Ho-based FiM alloys which have a stronger PMA due to the large orbital moment of the RE. The latter category has the potential to offer a higher stability of stored information, but the reports on skyrmions in these systems are scarce~\cite{chen2020observation}.

By contrast, for weak PMA FiM alloys, like FeGdCo, magnetic skyrmions have been recently observed~\cite{kim2019bulk} and they can be controlled in microstructured devices~\cite{woo2018current,wu2020ferrimagnetic}. Since then much research has been carried out to enable control and fuctionalization of ferrimagnetic skyrmions: they have been observed in ferrimagnetic confined nanostructures~\cite{brandau2019};  topological spin memory is reported for Co/Gd multilayers exhibiting skyrmion stability in fully compensated antiferromagnetically coupled heterostructure~\cite{wang:2022};observation of spin spirals and individual skyrmions in synthetic Pt/CoGd/Pt ferrimagnetic multilayers at room-temperature has been achieved without the assistance of external magnetic fields~\cite{brandao:2022}; magneto-transport measurements have revealed a topological contribution resulting from the occurrence of an interfacial DMI  in Ho/CoFeGd/$\beta-$W multilayers~\cite{ramesh:2022}; evidence for chiral ferrimagnetism in an  ultrathin GdCo  layer has been demonstrated through a combination of high-resolution Lorentz microscopy and XMCD~\cite{streubel:2018}; N{\'e}el-type homochirality has been observed over a large temperature range in Ta/Ir/Fe/GdFeCo/Pt multilayers using scanning electron microscopy with polarization analysis~\cite{seng:2021}; and information on N{\'e}el versus Bloch DWs can be inferred by a tilt geometry with Lorentz transmission microscopy as shown  for a Mn$_3$Sn topological antiferromagnet~\cite{xu:2021}. However, a direct determination of the type of the spin structures, namely N\'eel-type versus Bloch-type, was not experimentally reported, for neither of the two FiM classes.

An unambiguous determination of the skyrmion type is crucial for understanding the stabilization mechanism of skyrmions in these materials. Indeed, one would expect stabilization of N\'eel-type skyrmions in a thin film system with an interfacial DMI induced by engineering of the spin-orbit coupling of the ferrimagnetic layer with neighboring heavy-metal (HM) layers \cite{fert2013skyrmions}. However, this approach usually requires stacking ultrathin magnetic and HM layers into an asymmetric periodic multilayer in order to achieve a sizeable magnitude of DMI and, consequently, a small enough ($\sim100$~nm) skyrmion size \cite{moreau2016additive}. On the other hand, "bulk" DMI stabilizing chiral Bloch-type skyrmions requires an intrinsic lack of inversion symmetry within the material~\cite{bogdanov1994thermodynamically}. Alternatively, skyrmion bubbles having the same topological charge, but degenerate chirality can also be stabilized by dipolar interactions ~\cite{buttner2018theory}. Furthermore, dipolar interaction can compete with an interfacial DMI and change the spin rotation sense from N\'eel to Bloch type, or give rise to hybrid spin textures~\cite{legrand2018hybrid}. Therefore, unveiling the skyrmion type in FiM will shed light onto the microscopic spin Hamiltonian in this class of thin-film FiM alloys.

In this study, we report real space imaging of the magnetic structures in a FiM DyCo$_{3}$ thin film by means of scanning transmission X-ray microscopy (STXM) utilizing both x-ray magnetic circular dichroism (XMCD) and x-ray magnetic linear dichroism (XMLD) contrast~\cite{thole1985strong,kuiper1993x,kortright2000magnetization,schutz1987absorption,Luo2019}. Using XMCD-STXM, we directly observe well-isolated FiM skyrmions and their transformation to maze-like domains as a function of the out-of-plane external field. With XMLD-STXM, we demonstrate that these FiM skyrmions are N\'eel-type and the maze-like domains also show a preference of N\'eel-type domain walls. We confirm our experimental results to be consistent with  micromagnetic simulations. Please note that the experiments reported here are performed at low temperatures (26~K) which correspond to a  non-fully compensated magnetization state. The fully compensated magnetization for this ferrimagnet occurs at a temperature that is  well above the room temperature, therefore a vanishing skyrmion Hall effect is not addressed(see Supplementary Note S1.2).

\section*{Results and discussion}
\subsection*{Field dependence of Skyrmion size}

\begin{figure}[!ht]
	\centering
 \includegraphics[width=\textwidth]{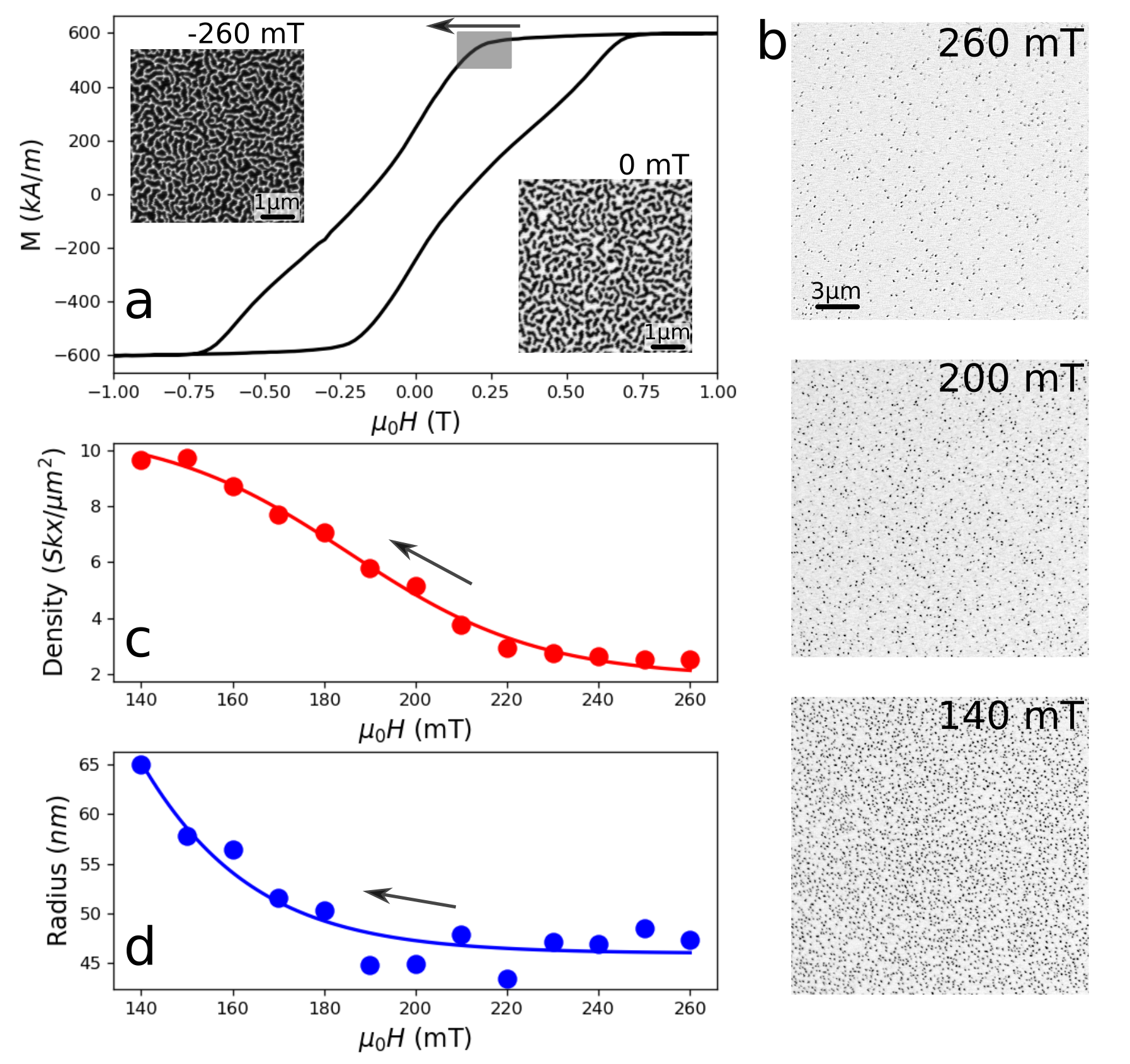}
	\caption{\figuretitle{ Hysteresis loop (measured by superconducting quantum interference device (SQUID) magnetometry) and scanning transmission X-ray microscopy (STXM) images of ferrimagnetic (FiM) skyrmions at 26~K.} (a) Out-of-plane magnetic hysteresis loop measured at 26~K by SQUID magnetometry. The external magnetic field $\mu_0 H$ is expressed in units and sub-units of Tesla(T), with $\mu_0$ being the vacuum magnetic permeability and $H$ denoting the magnetic field strength. (b) STXM images showing the FiM skyrmions at different perpendicular magnetic fields recorded at the Dy M$_{5}$ edge and 26~K. The average density (c) and radius (d) of the skyrmions as a function of external magnetic fields, were extracted from the STXM images. The grey rectangle box in panel (a) represents the magnetic field range where the FiM skyrmions are probed. The arrow represents the field sweeping direction. The insets of panel (a) show the transition into maze-like domains.} 
	\label{fig:hystloop}
\end{figure}

In Fig.~\ref{fig:hystloop}a we show the hysteresis loop in perpendicular geometry which was measured by SQUID magnetometry.
The magnetic reversal of the DyCo$_{3}$ film as a function of applied field was investigated at 26~K using STXM.  Because the maximum magnetic field for the STXM experiments is limited to 260~mT which is not enough to saturate the sample at low temperatures, we saturated the sample out-of-plane at room temperature prior to the STXM measurements. After cooling down the sample (in a perpendicular magnetic field of +260~mT) to 26~K, the STXM images were obtained with the magnetic field sweeping from +260~mT to -260~mT. 
Figure~\ref{fig:hystloop}b shows selected examples of XMCD-STXM areal images of the FiM skyrmions acquired at different perpendicular magnetic fields. One can distinguish the evolution of well-isolated FiM skyrmions as a function of  the magnetic field, ranging from 260~mT to 140~mT (within the field range that is marked by a grey rectangle in Fig.~\ref{fig:hystloop}a). The average density and radius of the skyrmions extracted from the STXM images as a function of field are shown in Fig.~\ref{fig:hystloop}c and in Fig.~\ref{fig:hystloop}d, respectively. It increases from 2.5 to 9.6  skyrmions per square micrometer when the field is decreased from 260~mT to 140~mT, indicating that the skyrmions can be created and annihilated by varying the field. At the same time, the average skyrmion radius $R$ increases from 45 nm to 65 nm, demonstrating that the skyrmion size can be reduced and inflated with the external field. When the magnetic field decreases further, the skyrmions will merge into worm- and maze-like domain structures, as shown in the insets of Fig.~\ref{fig:hystloop}a.

\subsection*{Lateral imaging of skyrmions with XMCD contrast}
\begin{figure}[t]
	\centering
	\includegraphics[width=\textwidth]{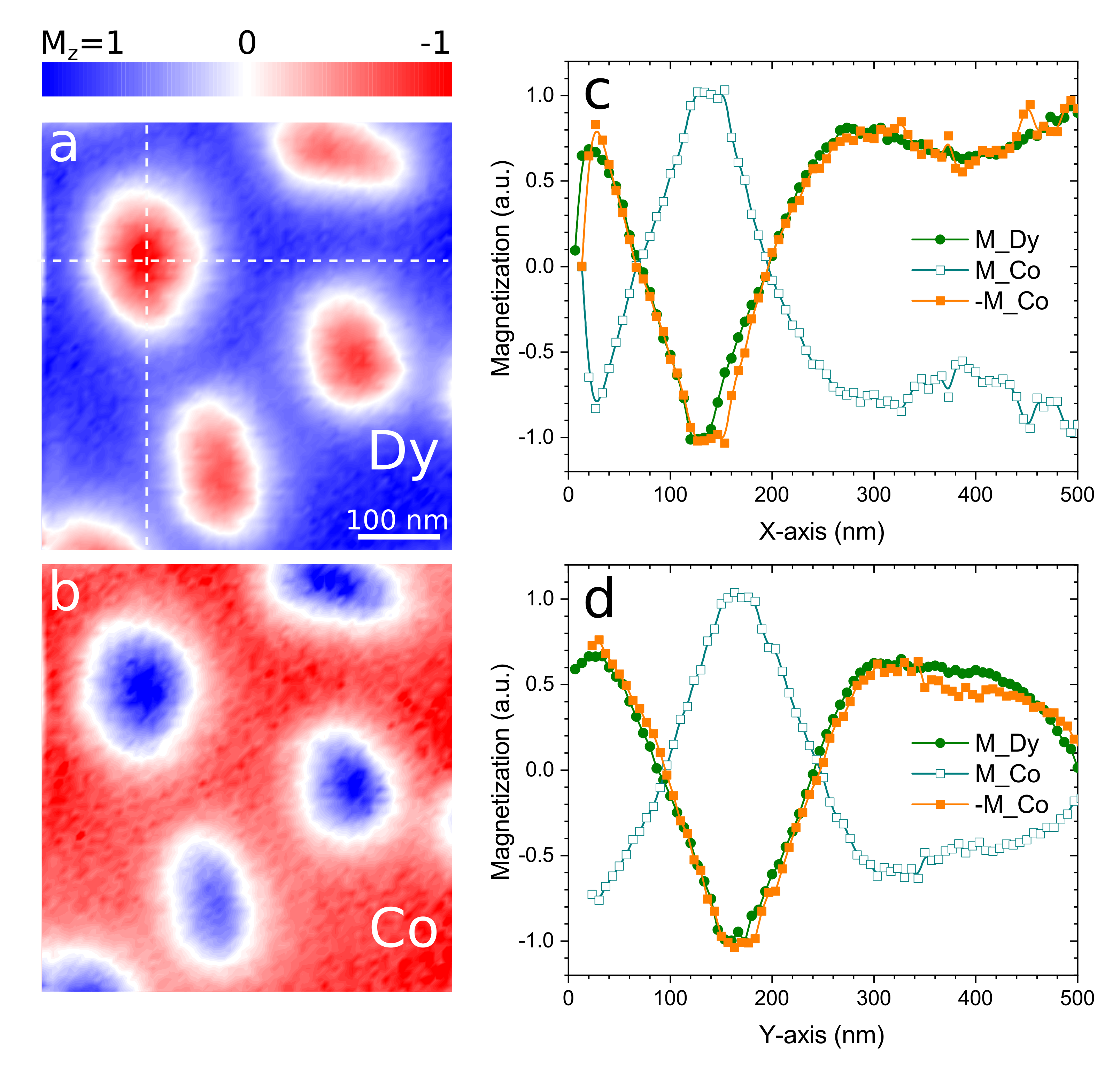}
	\caption{\figuretitle{X-ray magnetic circular dichroism imaging of isolated skyrmions.}(a) Scanning transmission x-ray microscopy (STXM) images at the Co L$_3$ edge and the Dy M$_5$ edge (b), respectively. The color bar represents the normalized sublattice magnetization ($\textrm{M}_\textrm{z}$) along the magnetic field (and the x-ray beam) direction.  (c, d) Line profiles across a skyrmion along the X and Y axes (see the dashed lines in panel a), representing the normalized perpendicular magnetic moment for Dy and Co elements.} 
	\label{fig:skyrmion}
\end{figure}

To resolve the details of the FiM skyrmions,  high resolution XMCD-STXM images were recorded at the Co L$_3$ and the Dy M$_5$ edges at an external field of 140~mT, as shown in Fig.~\ref{fig:skyrmion}. The left panels display the images of several single skyrmions, whereas in the right panels we show their line profiles along the X and Y axes, which represent the normalized perpendicular magnetic moment of Dy ($M_\textrm{Dy}$) and Co ($M_\textrm{Co}$) elements. One can easily observe that the magnetic profiles of $M_\textrm{Dy}$ and $M_\textrm{Co}$ overlap nicely, which clearly demonstrate the formation of FiM skyrmions with an antiparallel alignment of the Dy and Co moments. At this field, the radius (FWHM) of the respective skyrmion is about 65~nm.

\subsection*{Lateral imaging of skyrmion domain walls with XMLD contrast}

\begin{figure}[t]
	\centering
 \includegraphics[width=\textwidth]{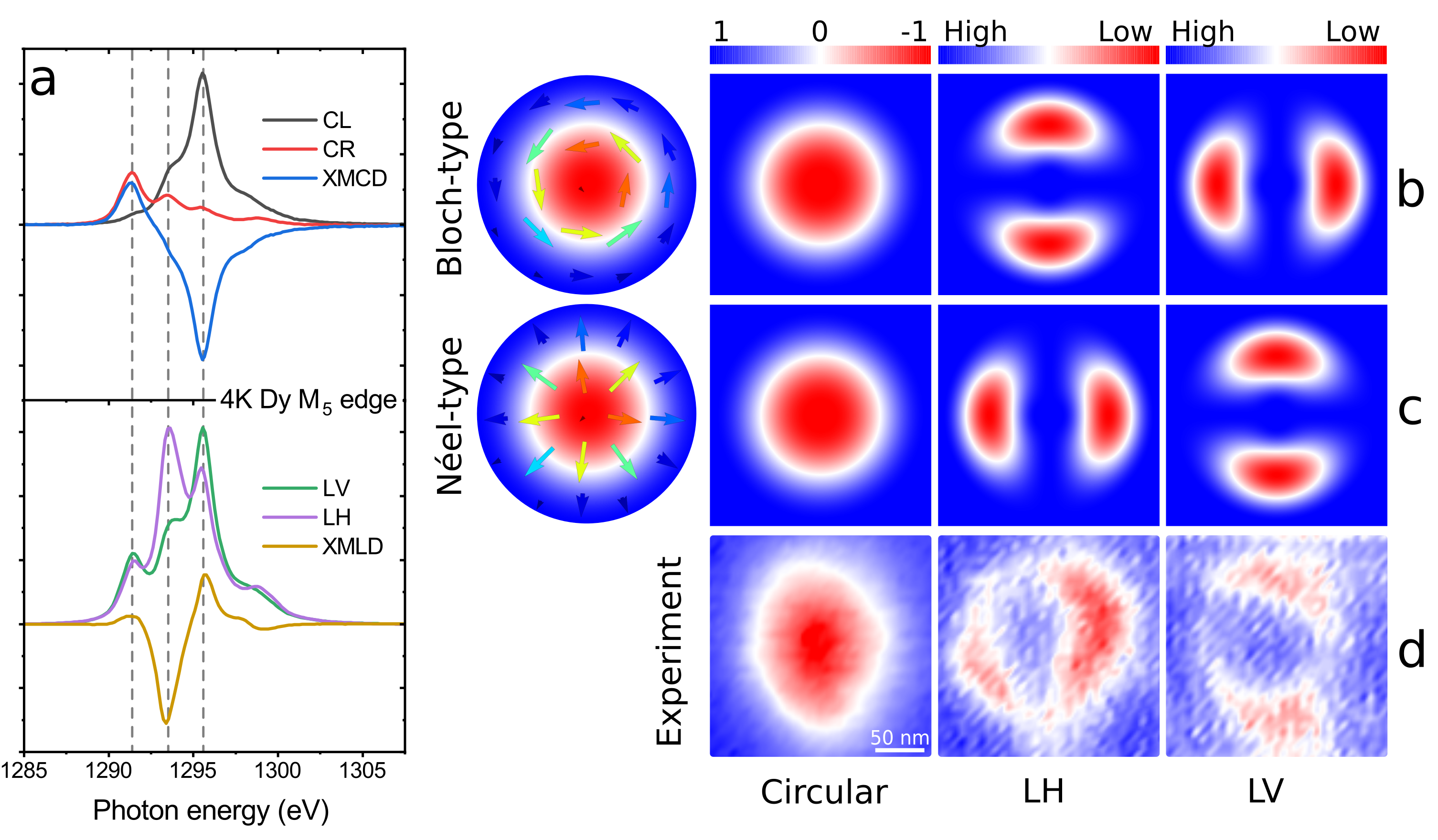}
	\caption{\figuretitle{Resolve of the magnetic domain wall type using x-ray magnetic linear  dichroism.}(a) x-ray absorption spectra (XAS) measured by circular left (CL), circular right (CR), linear horizontal (LH) and linear vertical (LV) polarized X-rays, as well as  x-ray magnetic linear dichroism (XMLD) and x-ray magnetic circular dichroism(XMCD) spectra at the Dy M$_5$ edge at 4~K. The three peaks of the Dy M$_5$ edge are marked by dashed lines. Expected magnetic image contrast when using circular, LH and LV polarized X-rays for Bloch-type (b) and N\'eel-type (c) skyrmions. (d) Experimental Scanning transmission X-ray microscopy (STXM) results. Here the STXM images for LV and LH x-ray polarizations were obtained at the middle peak of the Dy M$_5$ edge, and the STXM image for circular polarized X-rays was measured at the third peak of the Dy M$_5$ edge.} 
	\label{fig:XMLD}
\end{figure}

To identify the spin structure of the domain walls of the FiM skyrmions, XMLD was utilized by taking advantage of the strong linear dichroism of Dy at the M$_5$ absorption edge. Figure~\ref{fig:XMLD}a presents the demonstration of X-ray absorption spectroscopy (XAS) at the Dy M$_5$ edge for circular left (CL), circular right (CR), linear vertical (LV) and linear horizontal (LH) polarizations, respectively. The XMCD spectrum was obtained by taking the difference of (CR-CL), and the XMLD spectrum was obtained by taking the difference of (LV-LH). One can see that the maximum XMLD signal appears enhanced at the second peak of the Dy M$_5$ edge whereas the maximum XMCD signal is located at the third peak. The intensity of the XMLD spectra at the Dy edge is sufficiently large to be exploited for the STXM measurements~\cite{Luo2019}. Unlike XMCD which is sensitive to the magnetic moments collinear to the x-ray propagation direction, XMLD is sensitive to the magnetic moments collinear to the $\overrightarrow{E}$ vector of linearly polarized X-rays, which lies in the plane of the sample surface for our experimental geometry. To distinguish N\'eel-type and Bloch-type skyrmions from each other, we made simulations on how the two different types of skyrmions should look like in the presence of circular and linearly polarized X-rays, which are shown in Fig.~\ref{fig:XMLD}b and \ref{fig:XMLD}c. By comparison, one can see that our experimental data match very well with the N\'eel-type contrast, indicating that the FiM skyrmions in our DyCo$_3$ thin film are N\'eel-type skyrmions.

\subsection*{Lateral imaging of domain walls of a maze domain state with XMLD contrast}

\begin{figure}[t]
	\centering
  \includegraphics[width=\textwidth]{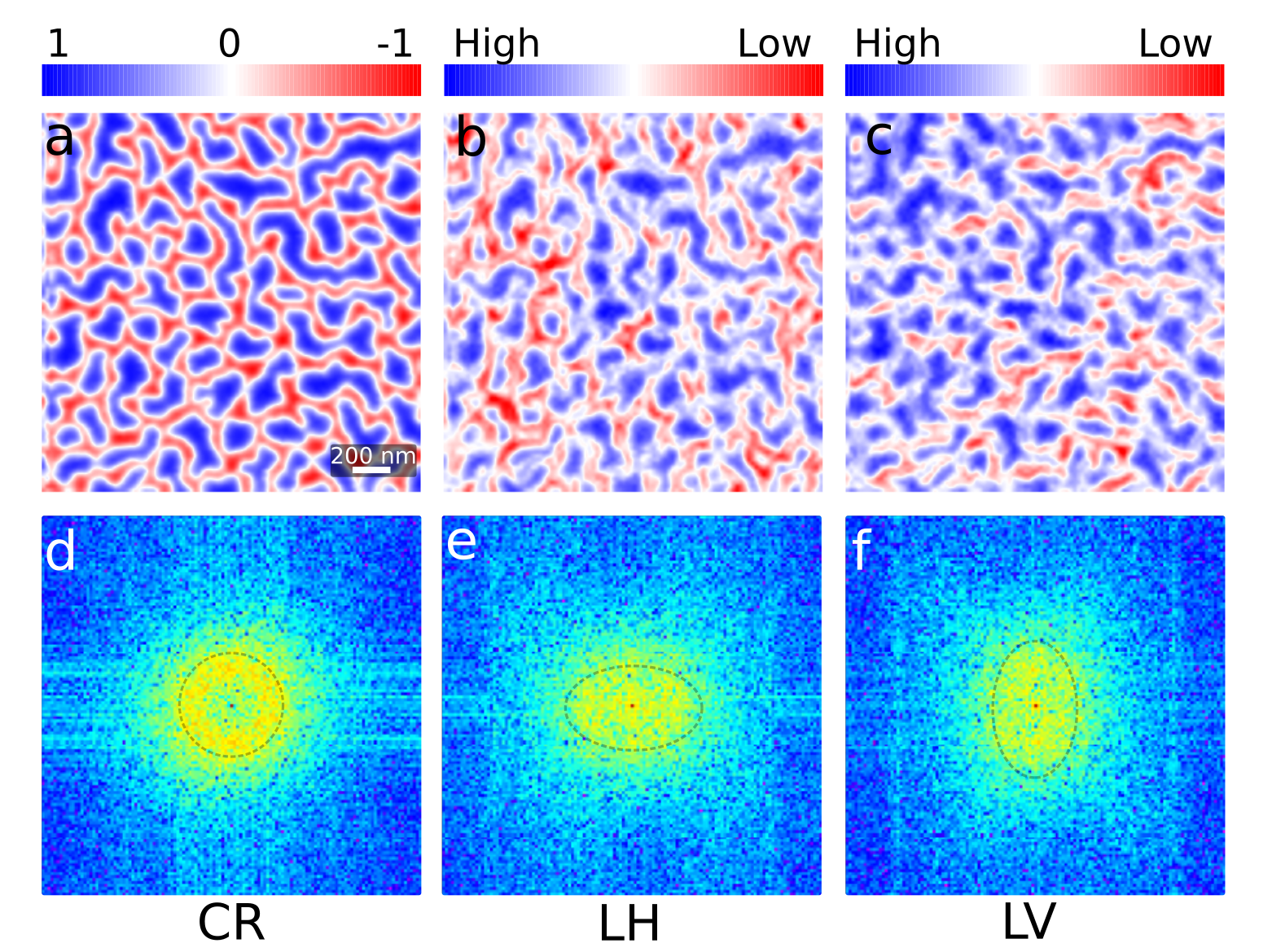}
	\caption{\figuretitle{Imaging of the magnetic domains and domain walls in a maze domain state and their corresponding Fast Fourier Transform.} (a-c) Scanning transmission X-ray microscopy (STXM) images of the maze domains at -260~mT for circular right (CR), linear horizontal (LH) and linear vertical (LV) polarized X-rays, respectively. (d-f) Fast Fourier Transform (FFT) of the top STXM images,  with the dashed circles and ellipses as guide for the eye.}
	\label{fig:maze}
\end{figure}

After successfully identifying N\'eel-type skyrmions utilizing the advantage of XMLD, we also investigated the domain walls for maze-like domains using the same technique, as shown in Fig.~\ref{fig:maze}(a-c). One can easily observe that the domain walls at the top/bottom sides are more pronounced for LV, and that the domain walls at the left/right sides show more intensity for LH, similar to the domain walls in skyrmions shown in Fig.~\ref{fig:XMLD}d. Note that, one still can see a weak contrast for some domain walls which do not follow this rule. This result indicates that the majority of the domain walls for maze-like domains are N\'eel-type with a low mixing of Bloch-type. We also applied Fast Fourier Transform (FFT) to the STXM images, which show different patterns for different polarizations (see Fig.~\ref{fig:maze}(d-f). The FFT shows a ring pattern for circular polarization. For linear polarization, however, it shows an ellipse with long vertical axis for LV and an ellipse with long horizontal axis for LH, which represent the preference of N\'eel-type domain walls (compare also to the results of micromagnetic simulations shown in the Supplementary Note S3 and Supplementary Note S4). 

\subsection*{Micromagnetic simulations}

Micromagnetic simulations were performed using the MuMax3 package~\cite{doi:10.1063/1.4899186} using  magnetic parameters for a DyCo$_3$ thin film deduced in a previous study \cite{chen2020observation} and in the present magnetometry measurements that can be found in the Supplementary Information file (see Supplementary Note S1.1).

Figure~\ref{fig:simulation} shows  simulated magnetic structures as a function of the magnetic field applied perpendicular to the sample plane. The top row shows the color-coded three-dimensional orientation of the magnetization, and the bottom row shows the in-plane component $M_x$. A maze domain pattern shows up upon a relaxation of the random magnetization state at zero field (Fig. \ref{fig:simulation}a), featuring also some N\'eel-type skyrmions with opposite polarities. Skyrmions with core magnetization parallel to the applied magnetic field collapse and disappear upon increasing the field (Fig. \ref{fig:simulation}b). Finally, at a higher field of $\mu_0 H =530$~mT the maze domain pattern evolves into a isolated skyrmion phase (Fig. \ref{fig:simulation}c). This phase persists up to $810$~mT when the skyrmions collapse and the sample gets fully magnetized along the field. Bottom panels of Figs. \ref{fig:simulation}a-d show the in-plane magnetization component within each cell of the simulation. The contrast inversion from blue (left) to red (right) of each stable N\'eel-type skyrmion represents the chirality of the system given by the sign of the DMI constant, which is not picked up by the STXM experiment. Importantly, the simulation captures accurately the impact of the domain wall type on the polarization-dependent STXM contrast. Circular and elliptic shapes of FFT patterns in Figs. \ref{fig:maze}d,e,f correspond very well to the ones calculated from the simulations (compare to Supplementary Figure S7) described in the Supplementary Note S3.

Interestingly, the simulated skyrmion size is very tunable and can be changed by the external magnetic field by a factor of three from $R=32$~nm at 530~mT to $R=9$~nm at 810~mT. While the skyrmion size differs from the experimental value quite significantly, lower DMI parameters do not allow to stabilize purely N\'eel-type skyrmions but rather hybrid ones that carry Bloch caps at the surface (see Supplementary Note S5.3), being consistent with the theory reported for magnetic multilayers \cite{lemesh2018twisted}. If no DMI is assumed, the interplay between PMA and the stray field results in the formation of a bubble lattice with dominantly Bloch-type domain walls (see Supplementary Note S5.1). Once a DMI term of sufficient strength is introduced, the stability of the skyrmions is increased towards higher fields, and the unique rotation sense of the domain walls gets defined. 
Note also that besides the DMI strength,  the saturation net magnetization and the magnetic anisotropy parameters play  an important role for the skyrmion formation in this material (see Supplementary Figure S3 and Supplementary Note S5.2).

For a discussion on the possible origin of a "bulk" DMI in this system see Ref.~\cite{chen2020observation} and the afferent Supplementary Note S6. Similar observations have recently been reported in Fe$_3$GeTe$_2$ flakes of various thickness where an interplay between dipolar and DM interactions results in a complex history-dependent magnetic phase diagram of spin textures \cite{birch2022history}.

\section*{Conclusions}
We presented an experimental resolve of skyrmion traits in a ferrimagnetic DyCo$_3$ thin film at 26~K using STXM imaging, utilizing both XMCD and XMLD contrast. With XMCD-STXM, the magnetic structures in real space are revealed as a function of decreasing external field. Well-isolated ferrimagntic skyrmions are observed between +260~mT and +140~mT, and the density as well as the radius of the skyrmions can be controlled by the external magnetic field. When the magnetic field is further reduced, these skyrmions will merge into maze-like domains, which matches very well with the results of  magnetic simulations. Utilizing XMLD-STXM at the Dy M$_5$ edge, we successfully identify the domain wall type of ferrimagnetic skyrmions to be of N\'eel-type. Moreover, the domain walls for the maze-like domains are also investigated, revealing also a majority of N\'eel-type domain walls. Hence, we are able to unambiguously conclude the  interfacial-type symmetry~\cite{SolrJuSER-juser_863478} of DMI in DyCo$_3$ thin film. Nevertheless, the origin of the strong DMI of this type remains an open question.
The technique of using XMLD contrast in the STXM measurements at the rare earth M edges provides a promising way to study complex spin textures in real-space, which is highly useful for the characterization of skyrmions, chiral domain walls and various non-collinear magnetic systems.

\section*{Methods}
\subsection*{Sample preparation and characterization}
The ferrimagnetic DyCo$_{3}$ film of 50~nm thickness was prepared by magnetron sputtering chamber (MAGSSY) at room temperature and in an argon atmosphere of $1.5\times10^{-3}$~mbar with a base pressure of $5\times10^{-9}$~mbar. The stoichiometry of the DyCo$_{3}$ alloy was controlled by varying the deposition rates of the Co and the Dy targets in a co-sputtering scheme. A Si$_{3}$N$_{4}$ membrane with a thickness of 100~nm was used as substrate for the soft X-ray transmission measurements. A capping layer of 3~nm thick Ta was deposited on top of the sample surface to prevent surface oxidation.
The magnetic properties of the sample have been measured by SQUID magnetometry and by anomalous Hall effect  (Tensormeter RTM1, HZDR Innovation, Germany), and they are described in the Supplementary Information file.

\subsection*{X-ray measurements}
Scanning transmission X-ray microscopy (STXM) measurements were performed at the MAXYMUS endstation at the Bessy II electron storage ring operated by the Helmholtz-Zentrum Berlin f\"{u}r Materialien und Energie~\cite{maxymus2012}. The X-ray beam was focused with a zone plate and an order selecting aperture on the transmissive sample in the presence of an applied out-of-plane magnetic field which was controlled by varying the arrangement of four permanent magnets. The STXM images were collected pixel by pixel using a piezoelectric sample stage at the Co L$_3$ edge and the Dy M$_5$ edge by exploiting the effects of x-ray magnetic circular  dichroism (XMCD) and x-ray magnetic linear  dichroism (XMLD). The XMLD contrast represents  an intensity map for LV (vertical axis in real space, parallel to the sample surface) and LH (horizontal axis in real space, parallel to the sample surface) orientations of the linear polarization axes. When the linear polarization is perpendicular to the spin axis, the XAS intensity measured at the middle resonance peak of the M5 edge is low (high in transmission), whereas for a parallel orientation of the linear polarization axis with respect to the spin axis the intensity is high (low in transmission). (see for instance Figure S1, in Ref~\cite{Luo2019}).  This makes the XMLD contrast easy to comprehend for the present transmission geometry of films with  perpendicular magnetic anisotropy: a change of intensity along the linear direction shown on the LV, LH and L45 maps (see Supplementary Note S2) can be given only by N\'{e}el walls, whereas a change of intensity towards a direction perpendicular to the linear polarization direction can be given only by Bloch walls. Note that the experimental XMLD maps shown in the manuscript are all logarithm of the raw data images.

The XAS, XMLD and XMCD spectra for the Dy M$_5$ edge (Fig.~\ref{fig:XMLD}a) were performed at the Deimos beamline at synchrotron Soleil~\cite{deimosbl2014} in transmission mode using the same 50~nm thick DyCo$_3$ sample grown on a Si$_3$N$_4$ membrane. The magnetic field of 2~T, which is much higher than the saturation field, was applied along the beam during the XMCD measurements using circular polarized X-rays and perpendicular to the beam during the XMLD measurements using linear horizontal and linear vertical polarized X-rays. 

\subsection*{Magnetic simulations}
Micromagnetic simulations were performed using MuMax3~\cite{doi:10.1063/1.4899186}. The simulation was performed on a three-dimensional grid $512\times512\times25$ voxels with the size of $2\times2\times2$~nm$^3$. The material parameters used for the simulation are given further below and can be found in the Supplementary Information file. A larger scale simulation with a grid of $1024\times1024\times25$ voxels was carried out for the simulation without  DMI, in order to account for the larger domain size. The computation was performed using a graphics processing unit (GPU) NVIDIA GeForce RTX 3080 Ti. The following material parameters were used: exchange stiffness $A_\textrm{ex}$~=~6~pJ~m$^{-1}$, saturation magnetization $M_s$~=~600~kA~m$^{-1}$, and uniaxial anisotropy K$^\textrm{u}$~=~130~kJ~m$^{-3}$. The interfacial-type DMI constant $D_\textrm{int}$ was tuned to obtain the isolated field-induced N\'eel-type skyrmion phase without admixtures of maze domains. The minimal value of DMI required for the purely N\'eel-type skyrmion stability was found to be $D_\textrm{int}$=0.0015~J~m$^{-2}$. It is remarkable that this value amounts about 8\% of the  exchange energy of DyCo$3$~\cite{yakintohos1975}, in  agreement with the suggestion of up to $\sim$20\% of the isotropic exchange expected for DMI in disordered systems~\cite{fert1980role,fert1990} (see Supplementary Note S6).




\section*{Acknowledgments}

We thank the Helmholtz-Zentrum Berlin für Materialien und Energie for the allocation of synchrotron radiation beamtime (Proposal No. 212-10386). The authors acknowledge  financial support  by the German Federal Ministry for Education and Research (BMBF project No. 05K19W061). F.R. acknowledges financial support by the German Research Foundation via Project No. SPP2137/RA 3570. We acknowledge the use of the Physical properties laboratory, which is part of the CoreLab "Quantum Materials" operated by HZB. F.R. acknowledges insightful information provided by Dr. Eugen Weschke on the UE46 undulator operation.

\section*{Contributions}
C.L., K.C. and F.R. conceived and designed the experiments. K.C. prepared the samples. C.L. and F.R. performed the STXM experiments with the help of S.W. and M.W., C.L. and K.C. analyzed the STXM data and prepared the figures. V.U. performed the micromagnetic simulations. K.P. performed the magnetic characterization by SQUID. R.A., C.L., V.U. and F.R. performed the magneto-transport experiments. C.L., V.U. and F.R. wrote the manuscript draft. All authors discussed the results and contributed to the manuscript.

 \clearpage

\begin{figure}[t]
	\centering		
	\includegraphics[width=\textwidth]{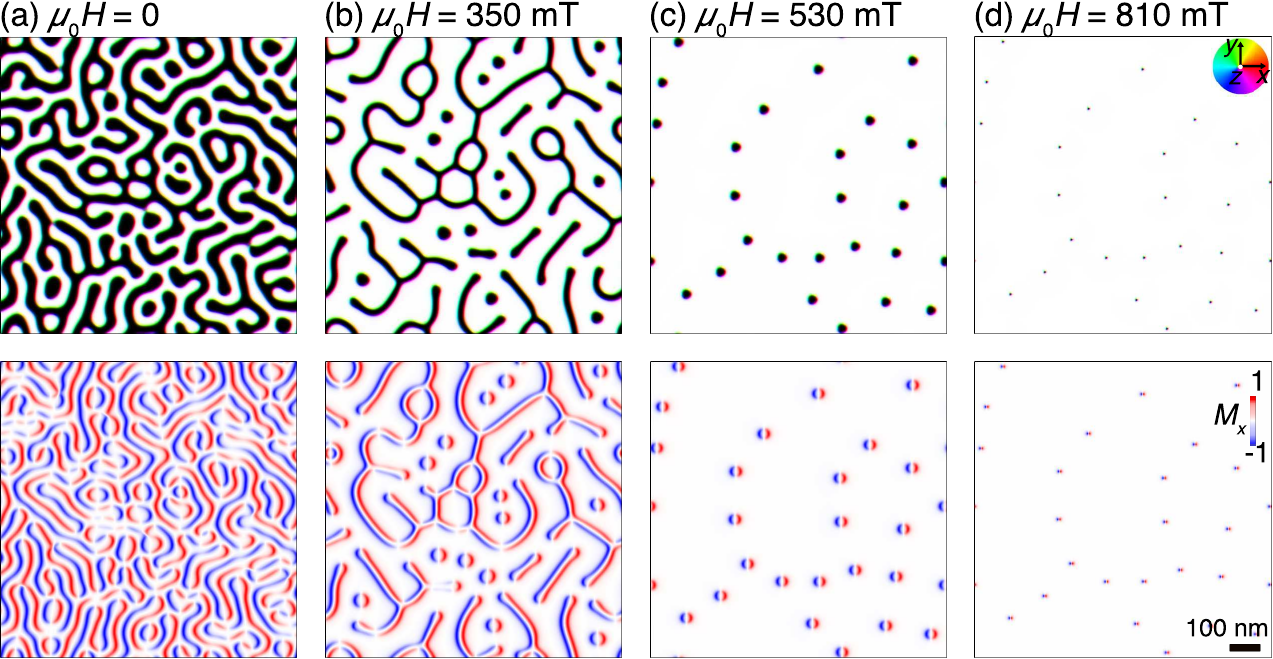}
	\caption{\figuretitle{Micromagnetic simulations demonstrating N\'eel-type skyrmion formation in the presence of a finite Dzyaloshinskii–Moriya interaction.} Top row of (a,b,c,d) panels: Micromagnetic simulations of relaxed spin configurations for the DyCo$_{3}$ film as a function of out-of-plane magnetic field. The external magnetic field $\mu_0 H$ is expressed in sub-units of Tesla(T), with $\mu_0$ being the vacuum magnetic permeability and $H$ denoting the magnetic field strength. The white and black colors represent the net normalized  magnetization being parallel and antiparallel to the $z$-axis, respectively. The other colors  represents the orientation of the in-plane component of the net magnetization within each cell as shown in the color wheel in top panel (d). Bottom row of (a,b,c,d) panels: the   magnetization component along the $x$-axis. The  color bar shown in the bottom panel (d) represents the normalized net magnetization  $M_\textrm{x}$.}
	\label{fig:simulation}
\end{figure}

\clearpage
\renewcommand{\thefigure}{S\arabic{figure}}
\setcounter{figure}{0}
\renewcommand{\thefigure}{S\arabic{figure}}
\renewcommand{\thesection}{S\arabic{section}}
\renewcommand{\thesubsection}{S\arabic{section}.\arabic{subsection}}
\setcounter{section}{0}
\clearpage

\section*{Supplementary Information}
\section{Magnetic Characterization}
\subsection{SQUID magnetometry}
The magnetic properties of the sample have been measured by SQUID magnetometry. The sample has been cooled down to 2~K in a magnetic field of 2~Tesla and magnetic hysteresis loops have been measured as a function of temperature, on warming. They are shown in the Supplementary Figure~\ref{fig:figs1}. The left axis displays the magnetization in $\textrm{kA/m}$ as a function of an external magnetic  field which was applied in a direction perpendicular to the sample surface. At each temperature a field dependent magnetization was measured without the sample to correct for the eventual contributions from the sample holder itself. The absolute value of the magnetization is obtained by dividing the corrected SQUID raw response (emu) on the volume of the layer which is area $\times$ thickness, where the measured surface area was measured to be $6.53 \times 10^{-6} \textrm{m}^2$ and the thickness of the film is $50 \times 10^{-9} \textrm{m} $. The hysteresis loops exhibit a typical characteristic shape for films with perpendicular magnetic anisotropy, showing the onset of the magnetization reversal at the nucleation field (the magnetic field where the global remagnetization initiates), followed by magnetic domains formation down to the annihilation field where the magnetization is fully reversed.

The nucleation field is extracted for each temperature and is shown in  Figure~\ref{fig:figs2}. It exhibits a peculiar behavior: it has negative values at room temperature, changes sign at an intermediate temperature of about 230 K, increases in absolute values up to about 50 K and decreases again, even to negative values, towards lower temperatures. This type of behavior is expected for ferrimagnetic alloys due to the interplay between the anisotropy and the demagnetization energies.

in Supplementary Figure~\ref{fig:figs3} the saturation net magnetization extracted at the highest applied field is shown together with the magnetization at the nucleation field. The net magnetization increases from room temperature towards  lower temperatures in a monotonic way. Interestingly, the net magnetization at the nucleation field begins to deviate significantly from the saturation magnetization at a temperature of about 150 K. This temperature range (150~K to 2~K) can be considered as the phase diagram for the  skyrmions formation.

With the magnetic parameters determined from the hysteresis loops, one can extract the uniaxial magnetic anisotropy of the film as~\cite{aharoni2000introduction,carcia:1985}:
\begin{equation}
K^u=\frac{1}{2}  \mu M_{net} (-H_n+4 \pi M_{net})
\end{equation}
where $H_N$ is the nucleation field, $M_{net}$ is the net magnetization at the nucleation field. The results are shown in Supplementary Figure~\ref{fig:figs4}. We observe that the magnetic anisotropy changes as function of temperature in an expected fashion. For  ferrimagnetic films, it  is dominated by the single ion anisotropy which further follows the character of the orbital magnetic moment of the rare earth. The orbital magnetic  moment was measured previously by soft x-ray spectroscopy~\cite{PhysRevB.91.024409}. It exhibits a non-monotonic increase below 200~K which correlates well with the temperature dependence of the anisotropy determined from the SQUID data.


\begin{figure}
	\centering
	\includegraphics[width=1\linewidth]{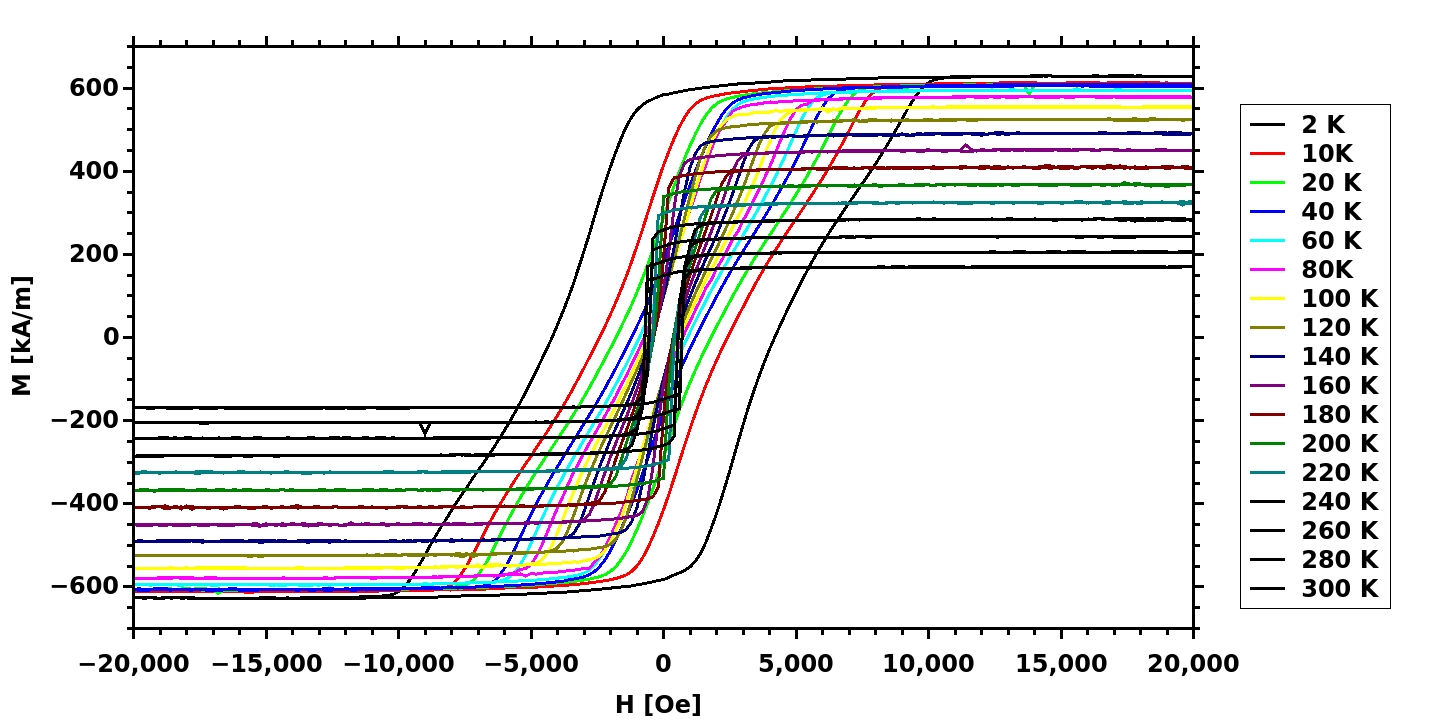}
	\caption{Hysteresis loops measured by SQUID at different temperatures with field applied perpendicular to the sample's surface.}
	\label{fig:figs1}
\end{figure}

\begin{figure}
	\centering
	\includegraphics[width=1\linewidth]{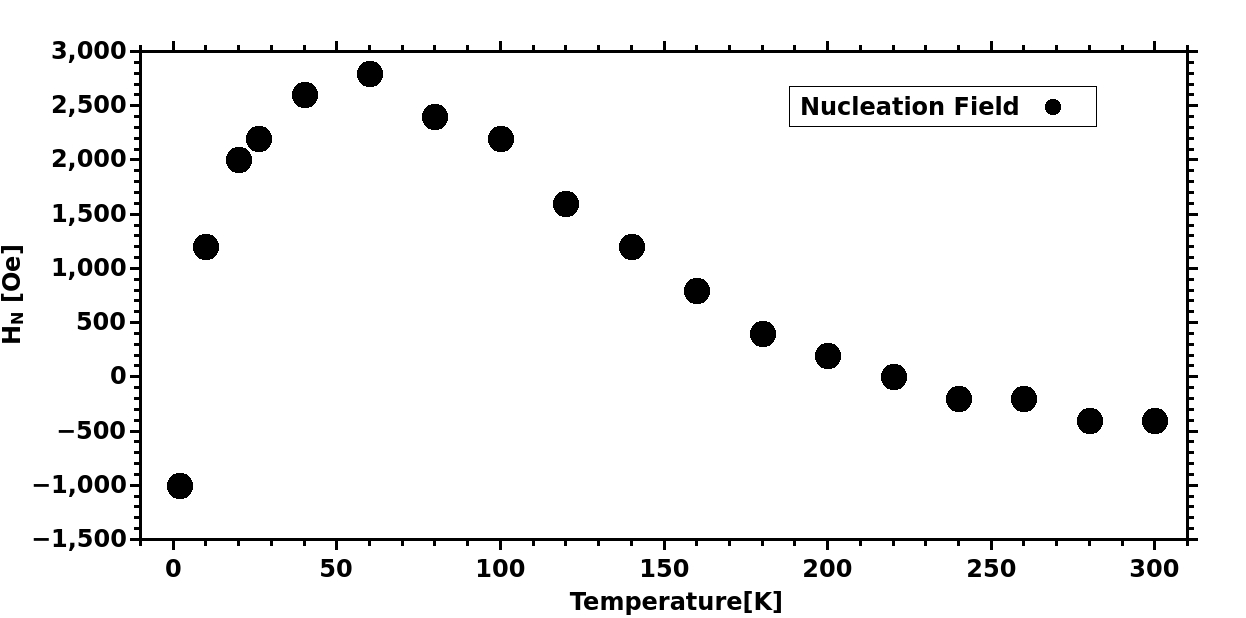}
	\caption{Temperature dependence of the nucleation field determined from data shown in Supplementary Figure~\ref{fig:figs1} }
	\label{fig:figs2}
\end{figure}

\begin{figure}
	\centering
	\includegraphics[width=1\linewidth]{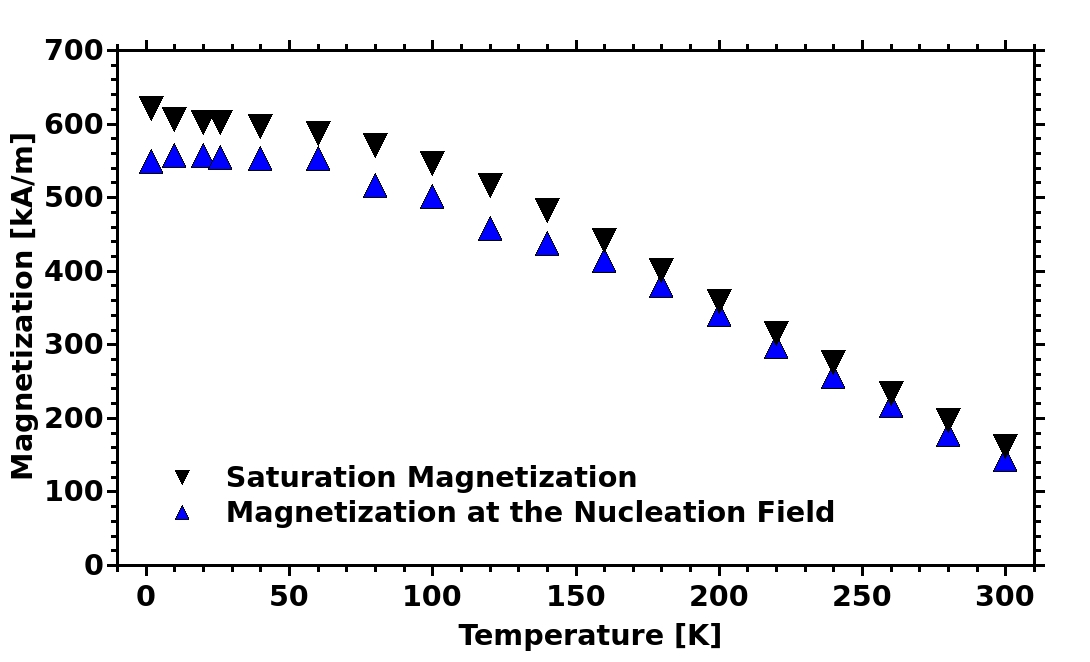}
	\includegraphics[width=1\linewidth]{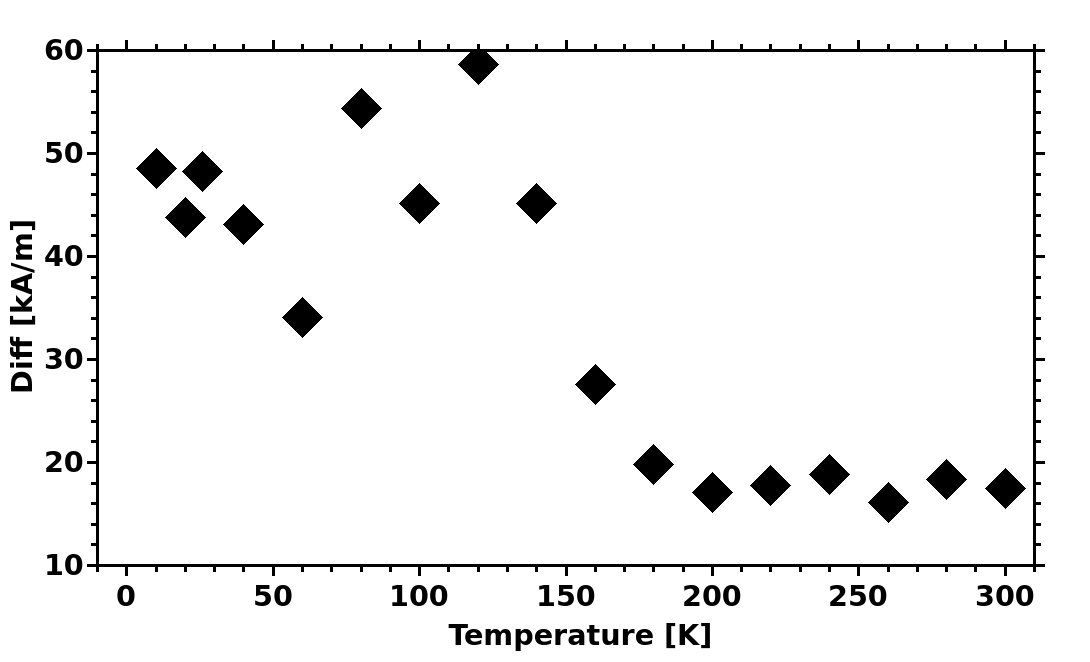}
	\caption{Top panel: Saturation net magnetization extracted from the hysteresis loops at 2~Tesla (black down-triangles) and the magnetization at the nucleation field(blue up-triangles). Bottom panel: The difference between the saturation net magnetization and the magnetization at the nucleation field. This panel can be considered at a phase diagram for temperature range (0-150K) of skyrmions formation.}
	\label{fig:figs3}
\end{figure}

\begin{figure}
	\centering
	\includegraphics[width=1\linewidth]{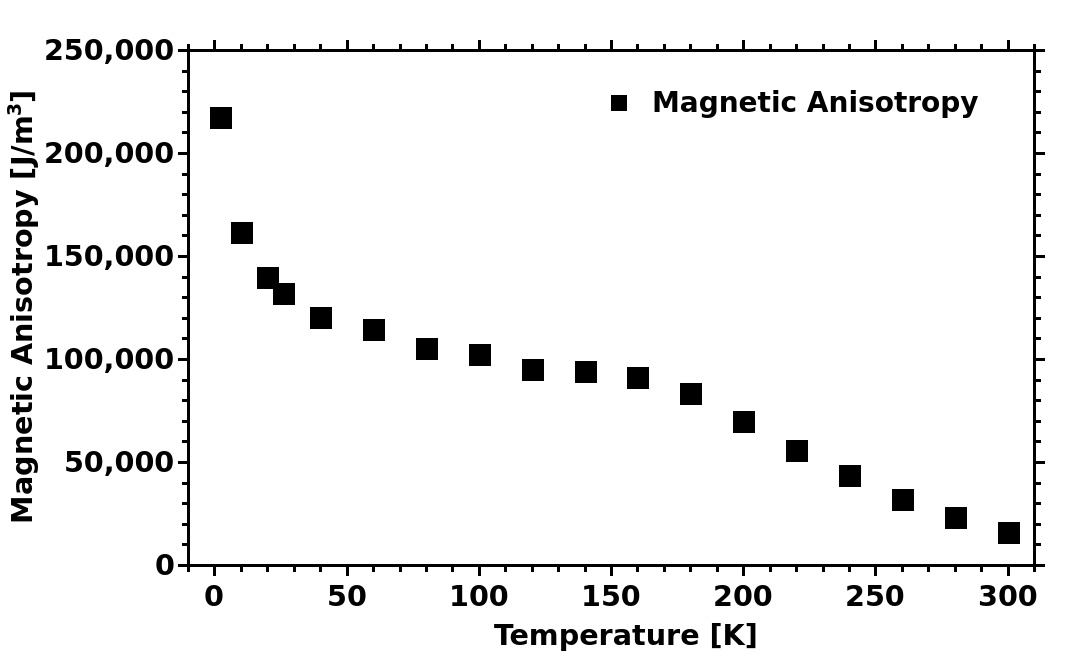}
	\caption{Temperature dependence of the magnetic anisotropy calculated according to Supplementary Equation~1.}
	\label{fig:figs4}
\end{figure}
 \clearpage

\subsection{Electrical Transport}

The characterization across the compensation magnetization was performed by magneto-transport measurements as a function of  temperature from 290~K to 460~K. The method involved here is spinning-current  anomalous  Hall  magnetometry~\cite{kosub:prl:2015} which is implemented in a commercial device  named Tensormeter (HZDR Innovation, Germany). This spinning-current approach has the advantage that extrinsic parasitic contributions to the anomalous Hall output signal are compensated dynamically which lifts the need of sample microstructuring.

The sample has been wire-bonded and connected with low resistance wires to the measuring device. The transport measurements were performed under high vacuum($3\times 10^{-8}$~mbar) with the Alice II instrument~\cite{lit:ALICE-RSI,ukleevSTAM2022}.  The Hall resistance R$_{xy}$ of the sample was measured in a four-wire configuration using the Zero-Offset Hall preset of the Tensormeter. The temperature was calibrated by a temperature sensor mounted on the sample holder and the magnetic field was applied perpendicular to the sample's surface. 

In Supplementary Figure~\ref{fig:tcomp}a we show few selected hysteresis loops at temperatures below and above the magnetization compensation. The hysteresis loops show an increased coercive field close to the compensation temperature. Moreover, the  sign of the hysteresis loops reverses as the temperature increases above the compensation. This effect reflects the sensitivity of the anomalous Hall resistance (R$_{xy}$) which \textit{in effectum} is proportional to  the difference of the up and down electronic density of states at the Fermi energy. For 3d ferromagnets  this is proportional to the net magnetization. However,  for  RE-TM ferrimagnets alloys the 3d density of states at the Fermi energy are dominated by the TM element (in our case Co), whereas the density of 4f-states of the RE ion (in our case Dy) are more localized below the Fermi energy(see Fig. 1c of \cite{abrudan:pss:2021}), therefore TM is contributing less significantly to the anomalous Hall resistivity. 

The coercive field of each hysterysis loop was extracted and plotted in the Supplementary Figure~\ref{fig:tcomp}b. At the divergence position, the net magnetization of the film vanishes. This corresponds to magnetic compensation temperature $T_{comp}$, which is for this sample 415~$\pm$~5 K.

\begin{figure}[h]
	\centering
	\includegraphics[width=1\linewidth]{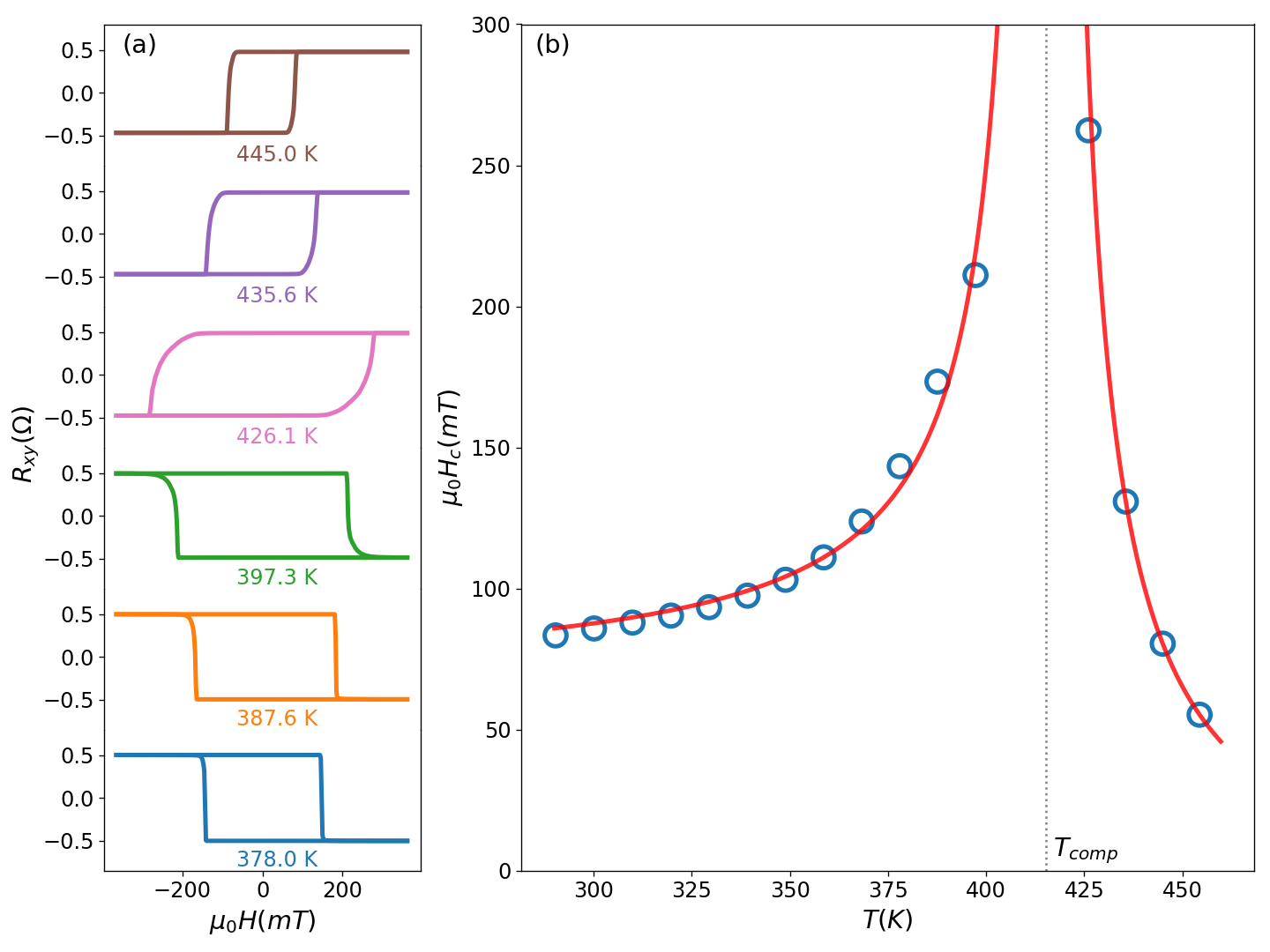}
	\caption{The anomalous Hall effect (AHE) was measured at different temperatures. (a) The Hall resistance R$_{xy}$ as a function of perpendicular magnetic field. The hysteresis loops change sign between 397.3 K and 426.1 K, which indicates the crossing of the magnetic compensation temperature $T_{comp}$. (b) Temperature dependence of the coercivity field $\mu_0 H_c$ exhibits a divergent behavior near $T_{comp}$ at $\approx$ 415~K.}
	\label{fig:tcomp}
\end{figure}

 \clearpage
 
\section{STXM images for a 45 degrees orientation of the linear polarization}

A  45\degree~rotation of the linear X-ray polarization  was implemented to investigate the skyrmions for larger areas, as demonstratively shown in Supplementary Figure~\ref{fig:L45}. By comparing such a STXM image to those with LH and LV polarization, it can be seen that the contrast of the domain walls also rotates by 45 degrees, which further confirms that the skyrmions are N\'eel-type.

Note that in spite of a clear N\'eel character, the the skyrmion and domain walls  shape  exhibit distortions. Given that the atomic arrangements in amorphous alloys are not well-defined,  local variation of magnetic anisotropy and stiffness may impact on the homogeneity of the magnetic ground states. Also,  local  pinning at eventual defect sites may contribute to complex skyrmion shape distortions~\cite{Gruber2022}.

\begin{figure}[h]
	\centering
 \includegraphics[width=\textwidth]{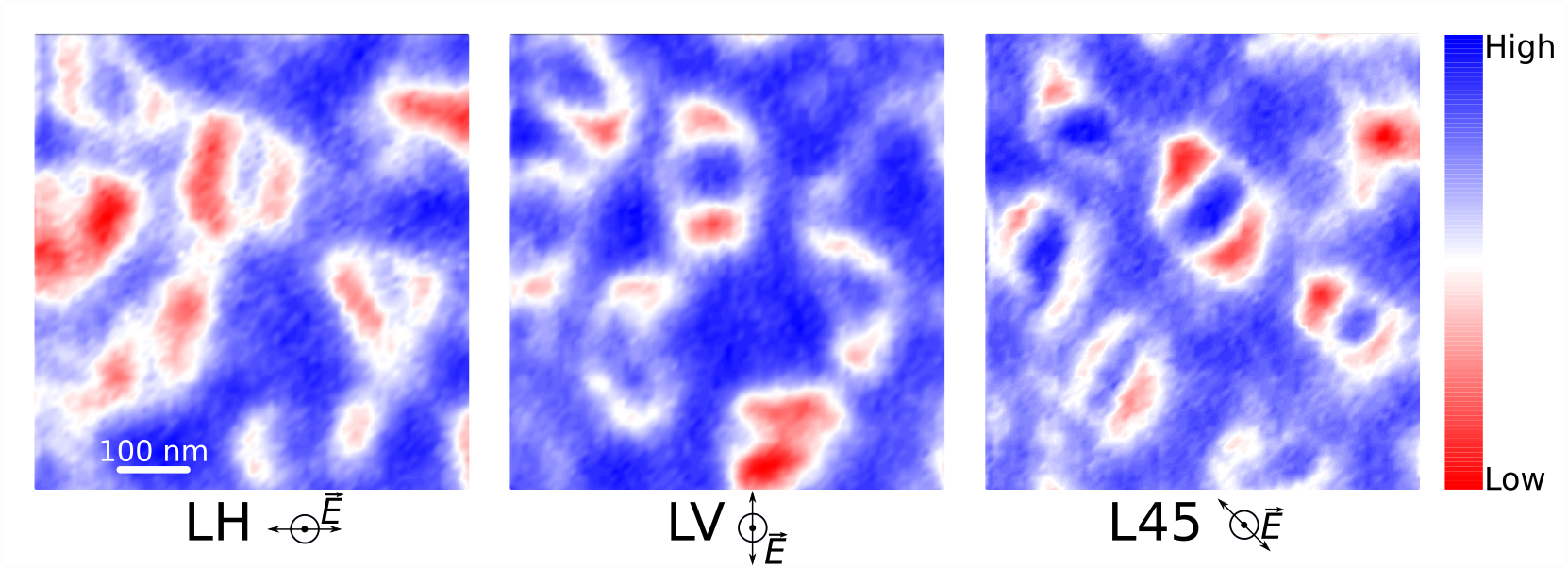}
	\caption{ XMLD-STXM images of skyrmions at 140~mT and 30~K for LH and LV, and 45\degree ~linear x-ray polarization.} 
	\label{fig:L45}
\end{figure}

\section{Micromagnetic simulations: Fast Fourier Transform of the maze-domain state with N{\'e}el walls}

The $M_z$, $M_x$, and $M_y$ magnetization components of the maze domain pattern resulting from the micromagnetic simulations are shown in Supplementary Figure~\ref{fig:sim_fft}, together with  their Fast Fourier transform (FFT) patterns. The FFTs correspond well to the experimental data shown in Figure~4(d-f) of the main manuscript. While the FFT of the $M_z$ component, which corresponds to the contrast mechanism to the circular light, results in a ring of intensity (Supplementary Figure~\ref{fig:sim_fft}a), the FFT patterns for the square of $M_x$, $M_y$ show arcs of intensity aligned in horizontal and vertical directions, respectively (Supplementary Figure~\ref{fig:sim_fft}b,c). On the other hand, the FFTs of the experimental data measured with linear horizontal and vertical light polarizations (Figure~4e,f)  appear as elongated ellipses because of a broader width distribution of the maze domains in the real sample, which is further convoluted  with the finite resolution of STXM images.

\begin{figure}[!h]
	\centering
	\includegraphics[width=\textwidth]{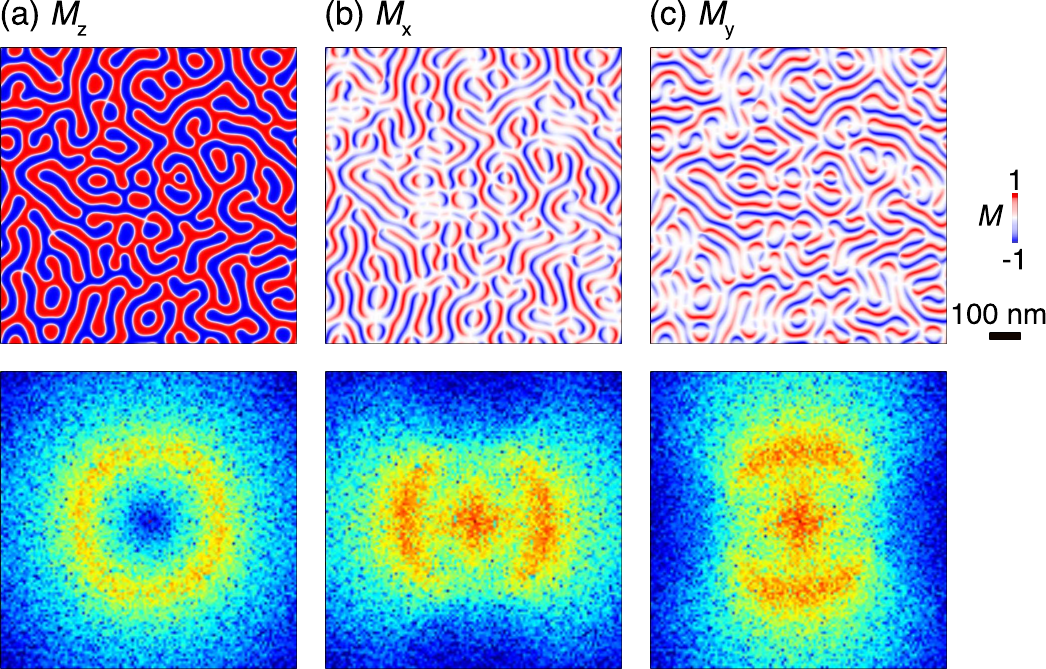}
	\caption{ Micromagnetic simulations of the zero-field maze domain pattern for the DyCo$_{3}$ film. Magnetization components (a) $M_z$, (b) $M_x$, and (c) $M_y$ are shown, respectively. First (from the left) bottom panel shows the Fast Fourier transform (FFT) of the $M_z$ component while the other two bottom panels show the FFT of the square of the corresponding upper $M_x$ and $M_y$ patterns.} 
	\label{fig:sim_fft}
\end{figure}

\section{Micromagnetic simulations: Fourier Transform of the maze-domain state with Bloch walls}

\begin{figure}[!ht]
	\centering
	\includegraphics[width=\textwidth]{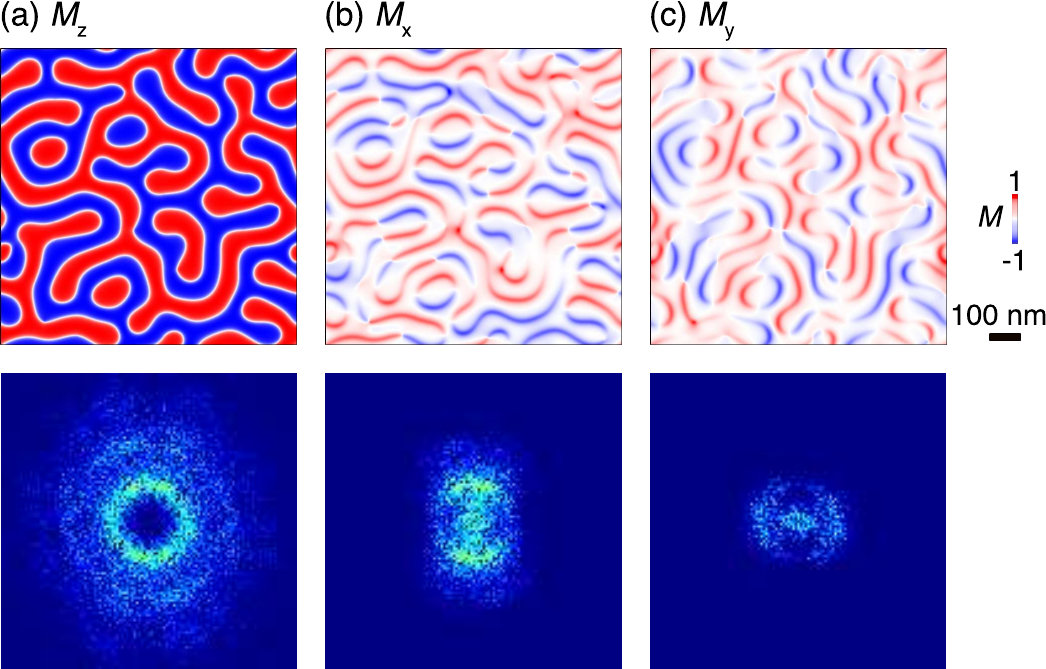}
	\caption{ Micromagnetic simulations of the zero-field maze domain pattern for the DyCo$_{3}$ film without taking DMI into account. Magnetization components (a) $M_z$, (b) $M_x$, and (c) $M_y$ are shown, respectively. First (from the left) bottom panel shows the Fast Fourier transform (FFT) of the $M_z$ component while the other two bottom panels show the FFT of the square of the corresponding upper $M_x$ and $M_y$ patterns.} 
	\label{fig:sim_fft_0DMI}
\end{figure}

FFT of $M_z$, $M_x$, $M_y$ components were also calculated for the micromagnetic simulations results without taking a  DMI into account (Supplementary Figure~\ref{fig:sim_fft_0DMI}). While the ring-like intensity is clearly observed in the FFT of the $M_z$ component, FFTs for $M_x$ and $M_y$ are clearly rotated by 90$^\circ$ compared to the previous case (Supplementary Figure~\ref{fig:sim_fft}). This clearly indicates the Bloch type character of the maze domain walls for a vanishing DMI in contrast to the N\'eel type case when the DMI is present.

\section{Micromagnetic simulations: Few more relevant showcases }
\subsection{Case 1: Simulations for magnetic parameters corresponding to 26~K, without DMI}

\begin{figure}[!ht]
	\centering
	\includegraphics[width=\textwidth]{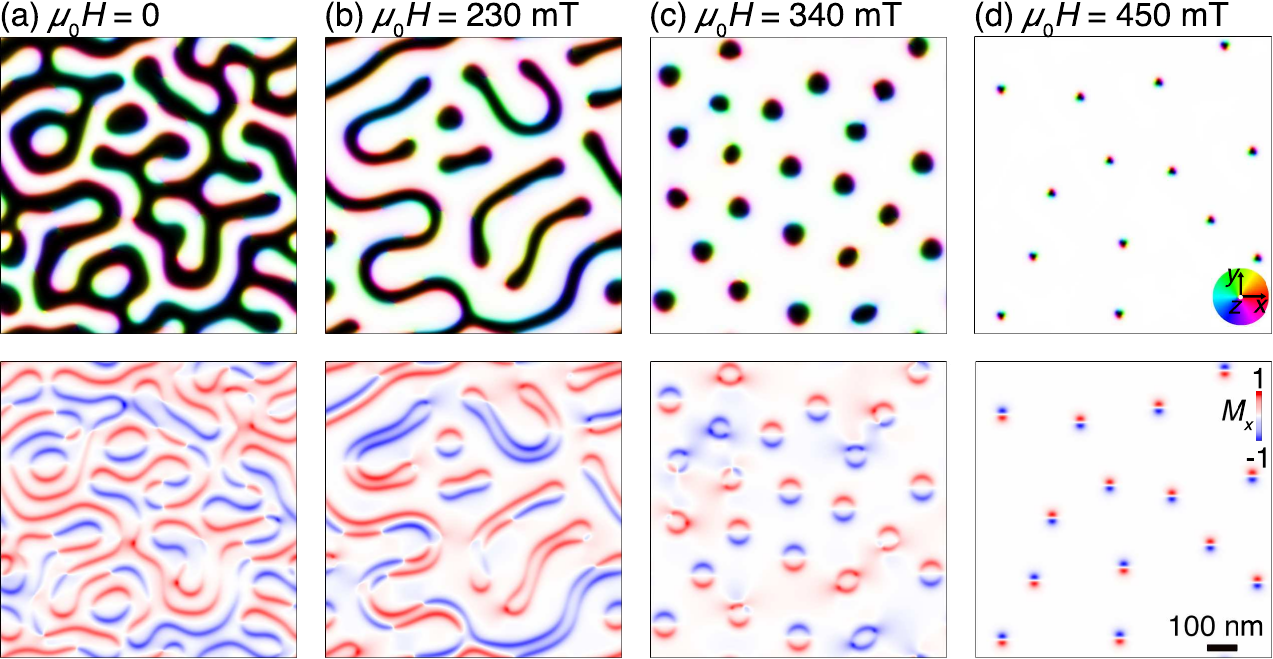}
	\caption{ Micromagnetic simulations of the spin configurations for the DyCo$_{3}$ film as a function of magnetic field without taking any DMI into account.The white and black represent the magnetization parallel and anti-parallel to the $z$-axis. The color code represents the orientation of the in-plane component of the magnetization within each cell as shown in the color wheel in the panel (d). The bottom panels show the magnetization component along the $x$-axis. The color code of the intensity of $M_\textrm{x}$ is given in the bottom panel (d).} 
	\label{fig:simulation2}
\end{figure}

The magnetic field dependence of the maze domain pattern has been calculated also for the case of a vanishing DMI, where the magnetic ground state  is determined by the interplay between dipolar interaction and uniaxial anisotropy only. Except for the DMI constant, all the other parameters in the simulation are similar to the ones in the main text. Supplementary Figure~\ref{fig:simulation2} shows the simulated magnetization distributions at zero applied magnetic field, 230\,mT, 340\,mT and 450\,mT. Relatively large ($\sim$100\,nm) magnetic domains are observed at zero field, that evolve into worm-like stripes and, finally, into skyrmion bubbles as higher out-of-plane magnetic field is applied. The size of the bubbles originally formed at 340\,mT decreases to 45\,nm at 450\,mT and then the skyrmion state collapses into a homogeneous one as the field is further increased. As a result, in the absence of DMI, maze domain walls and field-induced skyrmions are clearly identified as Bloch-type, which can be easily seen in the bottom panels of Supplementary Figure~\ref{fig:simulation2}, where the  intensity of the $M_x$ component changes from positive to negative at the top and bottom edges of the skyrmions. This is in  contrast to  the case of N\'eel type domain walls and skyrmions  where the intensity of these maps is rotated by 90 degrees (see main text).

\subsection{Case 2: Simulations for the magnetic parameters corresponding to 200~K with DMI}

\begin{figure}[!ht]
	\centering
	\includegraphics[width=\textwidth]{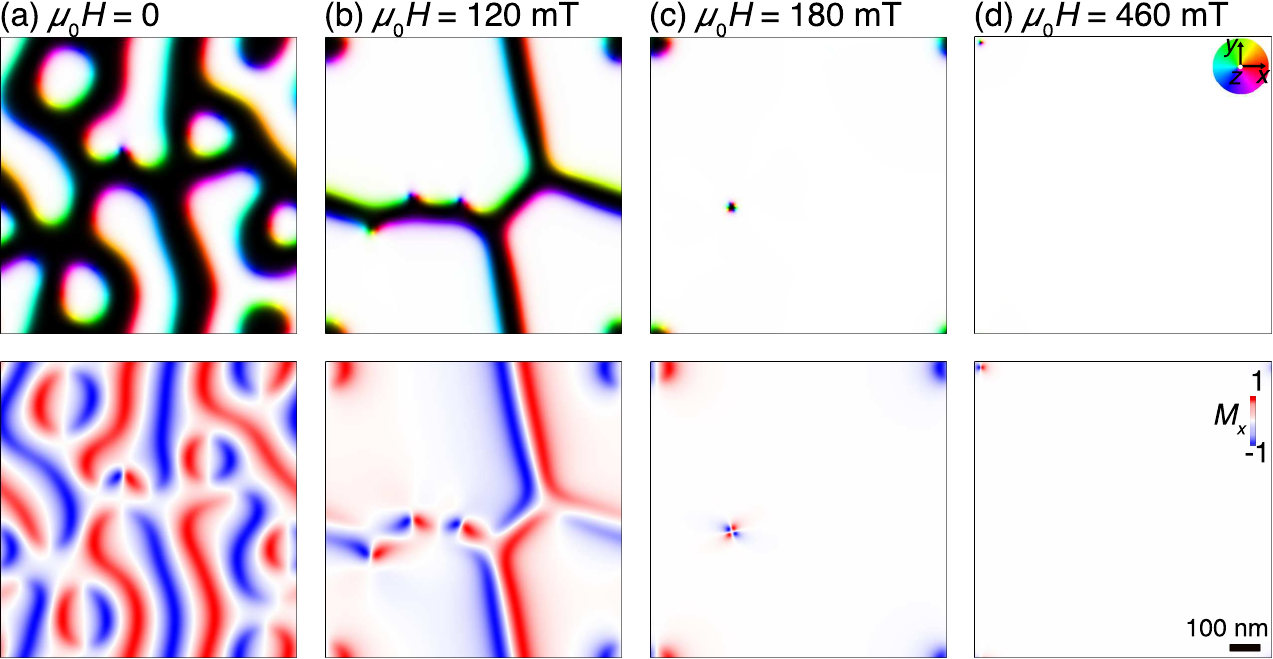}
	\caption{ Micromagnetic simulations of the spin configurations for the DyCo$_{3}$ film as a function of magnetic field with the $M_s$ and $K^\textrm{u}$ parameters corresponding to 200\,K. The black and white contrast indicates the magnetization being parallel and anti-parallel to the $z$-axis. The color code represents the orientation of the in-plane component of the magnetization within each cell as shown in the color wheel in the top panel (d). The bottom panels show the magnetization component along the $x$-axis. The color code of the intensity of $M_\textrm{x}$ is given in the bottom panel (d).} 
	\label{fig:simulation200K}
\end{figure}

Micromagnetic simulations for a saturation magnetization of $M_s=358.5$\,kA/m, an uniaxial anisotropy of  $K^\textrm{u}=69.8$\,kJm$^{-3}$ and DMI equal to $D_{int}=0.0015$\,Jm$^{-2}$ were carried out to mimic the sample behavior at 200\,K. The saturation net magnetization and magnetic anisotropy parameters were extracted from the magnetometry measurements. Here, large magnetic  domains (of the order of hundreds of nm) with N\'eel-type walls take place, evolving into a very sparse skyrmion state ($\sim 1$ per 1\,$\mu$m$^2$) at higher applied fields (Supplementary Figure~\ref{fig:simulation200K}). The simulation matches well with the STXM results which show similar large-scale textures at 200\,K \cite{chen2020observation} and absence of the skyrmion phase. We speculate that tuning the DyCo composition in order to increase the anisotropy and saturation magnetization towards room temperature can pave a way to stabilize technologically promising ferrimagnetic topological spin textures in DyCo at 300\,K. 

Interestingly, a single anti-skyrmion has been spotted at an intermediate field range of 150-180\,mT in the simulations (Supplementary Figure~\ref{fig:simulation200K}c), which is unexpected in the system with this type of DMI. Confirmation of this observation requires further computational and experimental studies.

\subsection{Case 3: Simulations for the magnetic parameters corresponding to an intermediate DMI parameter at 26K}

As we describe in the main text, the minimal value of the interfacial-type DMI constant required to stabilize N\'eel-type skyrmions in the micromagnetic simulations is found at $D_\textrm{int}=0.0015$\,Jm$^{-2}$, while lower values result in Bloch- or hybrid-type skyrmion textures. An example of such a hybrid skyrmion is shown in Supplementary Figure~\ref{fig:hybrid}. Here, the skyrmion winding changes from Bloch-type in the top layer of the film ($z=1$) to N\'eel-type in the bottom layer ($z=25$) via an intermediate state in the middle. This behavior correspond well to predictions given by Lemesh and Beach in their analytical theory for multilayers \cite{lemesh2018twisted}.  

\begin{figure}[!ht]
	\centering
	\includegraphics[width=\textwidth]{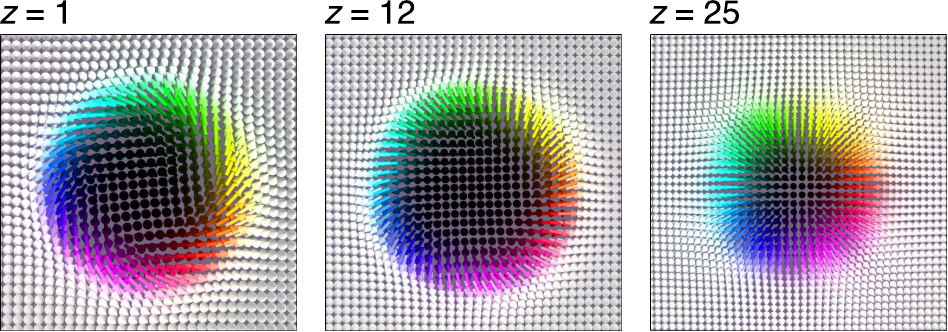}
	\caption{Magnetic structure of hybrid skyrmions in DyCo$_{3}$ film with an intermediate value of the DMI constant $D_\textrm{int}=0.00125$\,Jm$^{-2}$ at 470\,mT as obtained from the micromagnetic simulations. The panels show the single skyrmion structure in the top ($z=1$), middle ($z=12$) and bottom ($z=25$) layers of the film, respectively.} 
	\label{fig:hybrid}
\end{figure}
 \clearpage
\section{Possible origin of a "bulk" DMI in the ferrimagnetic alloys }
One main outcome of the resolve of the domain walls as N\'eel-type  is that they require a "bulk" DMI of an interfacial-type symmetry~\cite{SolrJuSER-juser_863478} to occur. In the absence of a DMI, the domain walls are Bloch-type as revealed by the micromagnetic simulations. It is  well known that  Ta or Pt buffer layers induce an interfacial DMI which has positive or negative sign, respectively. However, a DMI that is localized at the interface only is too weak to stabilize N\'eel walls. For instance, in FM/Ta bilayers, the iDMI has an noticeable impact only for a FM layer thickness that is below 1~nm~\cite{kopte2017complete}. In our case, the ferrimagnetic layer is 50~nm thick, and therefore a DMI localized at the interface only, is not likely to stabilize skyrmions as observed experimentally. Instead a bulk DMI owes to be responsible for the skyrmions formation in this system~\cite{chen2020observation}. \\ \\ The DyCo$_3$ single crystal has a complex rhombohedral structure where Dy occupies two sites, whereas Co occupies three sites in the unit cell. According to Ref.~\cite{YAKINTHOS1975979}, for one site the Dy  will align  itself  with the easy axis, whereas the Dy on the second site will prefer to align parallel to the c-axis of the crystal which is oriented perpendicular to the easy axis. As such, a non-collinear spin arrangement is possible due to the frustrated site magnetic anisotropy of the rare earth ions. For the amorphous  DyCo$_3$ there is also strong experimental evidence for the occurrence of a  noncollinear spin state.  In Ref.~\cite{coey1976} an amourphous DyCo3 thick film has been studied by Mössbauer spectroscopy. It has been observed that the Dy ion exhibit a sperimagnetic arrangement, whereas the Co sublattice is ferromagnetically ordered. The above  experimental evidence for noncollinear behavior  manifest intrinsically for both single crystal and amourphous DyCo$_3$. \\
\\
Certainly, non-collinear interactions are yet not sufficient to lead to  skyrmion formation. Two more aspects play an important role for the formation of skyrmions in our DyCo3 thin film, namely: an amorphous thin film naturally lacks a rotational and transnational symmetry, (i. e. a lack of inversion symmetry that is similar to  the B20 structures) which, according to Ref.~\cite{fert1990} may lead to a DMI that amounts up to about 10\% of the isotropic exchange;  and the dipolar interaction strength for thin ferrimagnetic films with perpendicular magnetic anisotropy varies as a function of temperature(see section S1). All these three preconditions concur  in our system on the formation of small isolated  N\'eel skyrmions and N\'eel walls in the maze domain state.\\
\\
The exchange energy for DyCo$_3$ reported in Ref.\cite{yakintohos1975} is  621~$cm^{-1}$. Using the conversion factor 1~meV~=~8~cm$^{-1}$ and considering an average lattice parameter equal to unit volume 0.52635 nm$^3$ to the power of 1/3, we obtain an estimated exchange energy per surface area equal to 0.019 $\textrm{J/m}^2$. It is remarkable the minimal value of the DMI required for N\'eel-type skyrmion stability determined through micromagnetic simulations (see the main texts) of 0.0015 $\textrm{J/m}^2$ is within the 20\% of the estimated isotropic exchange, in agreement with the suggestion in   Ref.~\cite{fert1980role,fert1990}.

\section*{References}
\bibliography{ref}
\end{document}